\documentclass[preprint,
aps,preprintnumbers,nofootinbib,eqsecnum,showpacs]{revtex4}

\usepackage{amsmath}
\usepackage{amssymb}
\usepackage{latexsym}
\usepackage{graphicx}
\usepackage[usenames]{color}

\newcommand{\Trb}{{\rm Tr_B}}
\newcommand{\Trf}{{\rm Tr_F}}

\begin{document}


\preprint{RIKEN-TH-96}


\title{%
Supersymmetric Field Theory Based on Generalized Uncertainty Principle}


\author{SHIBUSA Yuuichirou}
\affiliation{Theoretical Physics Laboratory, \\
RIKEN (The Institute of Physical and Chemical Research), Wako, 351-0198, Japan }
    
\begin{abstract}
We construct a quantum theory of free fermion field based on 
the deformed Heisenberg algebra 
$[\hat{x},\hat{p}]=i \hbar(1+\beta \hat{p}^2)$ where $\beta$ is 
a deformation parameter using supersymmetry 
as a guiding principle. 
A supersymmetric field theory with a real scalar field and a Majorana 
fermion field is given explicitly and we also find that the supersymmetry algebra is  
deformed from an usual one.  
\end{abstract}

\pacs{03.65.Ca, 03.70.+k, 11.10.Ef, 11.30.Pb}



\maketitle

\section{Introduction}
\label{sec:intro}

Physics in extremely high energy regions is particularly of interest to
particle physics. In particular, when we discuss gravity, it is
expected that there is a minimal length in principle.
String theory which has a characteristic scale $\sqrt{\alpha'}$, is 
one of the most successful theoretical frameworks 
which overcome the difficulty of ultra-violet divergence in quantum
theory of gravity.  
However, string theory has many difficulties in performing practical computations. 
Therefore if we construct a field theory which captures some 
stringy nature and/or includes stringy corrections, it would play 
a pivotal role in investigating physics in high energy regions 
even near the Planck scale. 
 
Some of the stringy corrections appear as $\alpha'$ corrections. 
In other words, it often takes the form as higher derivative
corrections i.e. higher order polynomial of momentum.
One way to discuss these corrections is deforming the 
Heisenberg uncertainty principle to a generalized uncertainty principle (GUP):
\begin{eqnarray}
\Delta \hat{x} \ge \frac{\hbar}{2\Delta \hat{p}} + \frac{\hbar \beta}{2}\Delta \hat{p},
\end{eqnarray}
where $\beta$ is a deforming parameter and corresponds to the square of the minimal length scale.  
If GUP is realized in a certain string theory context, $\beta$ would take a value of order the string scale ($\beta \sim \alpha'$). 
This relation comes from various types of studies such as on 
high energy or short distance behavior of 
strings \cite{Gross:1987kz}, \cite{Konishi:1989wk}, gedanken experiment 
of black hole \cite{Maggiore:1993rv}, de Sitter space
\cite{Snyder:1946qz}, the symmetry of massless particle
\cite{Chagas-Filho:2005at} and wave packets \cite{Bang:2006va}.

There are several canonical commutation algebra which lead to the GUP. 
Among these algebra we will focus on the algebra;
\begin{eqnarray}
[\hat{x},\hat{p}]=i\hbar (1+\beta \hat{p}^2).
\label{deformed alg}
\end{eqnarray}
This algebra is investigated in \cite{Kempf:1994su}-\cite{Hossenfelder:2005ed} and an attempt to
construct a field theory with minimal length scale is made in
\cite{Kempf:1996ss} by using the Bargmann-Fock representation in 1+1 dimensional spacetime.  
It has also been used in cosmology, especially in physics at an early universe 
(see for example, \cite{Ashoorioon:2004vm}-\cite{Ashoorioon:2005ep} and references therein).

In our previous paper \cite{Matsuo:2005fb}, we investigated the quantization of fields based
on the deformed algebra (\ref{deformed alg}) in the canonical formalism
in 1+1 dimensions and in the path integral formalism as well. Using
the path integral formalism we constructed a quantum theory of
scalar field in arbitrary spacetime dimensions. This theory has a non-locality
which stems from the existence of a minimal length. 

In this paper, we construct a quantum theory of free fermion field based on 
the deformed Heisenberg algebra. Where, we respect
supersymmetry as a guiding principle. This is because a string theory has this symmetry and 
we intend to construct a field theory which contains the stringy
corrections.  
Moreover, supersymmetry is also an useful tool to understand physics in 
ultra-violet momentum regions. It manages a behavior of system in
extremely high energy regions and eases ultra-violet divergence in quantum
theory. 
Therefore we propose a quantum field theory of fermion to have a
supersymmetry for a scalar system which was given in
\cite{Matsuo:2005fb}. 
In two and three-dimensional
spacetime, we give a system with one real scalar and one Majorana
fermion  explicitly. This system has 
a special symmetry between a boson and a fermion which corresponds to
supersymmetry. Although, this symmetry is deformed from ordinary
supersymmetry. From the fermionic part of this system, we propose an action 
of fermionic fields based on GUP in general dimensional spacetime.

\section{Scalar Field Theory}
\label{sec:scalar}
In the paper \cite{Matsuo:2005fb}, we proposed a field theory of
scalar based on GUP in the path-integral formalism. We begin with  
a review of this theory. 

Our theory is based on the following algebra \cite{Kempf:1994su}: 
\begin{eqnarray}
 \left[\hat{x}^i,\hat{p}_j\right]&=&i\hbar (1+\beta \mbox{\boldmath
  $\hat{p}$}^2)\delta^i_{\sp j}. \label{GUP2}
\end{eqnarray}
This is an  extension to higher dimensional spacetime of deformed Heisenberg
algebra (\ref{deformed alg}).
Here $i,j$ run from 1 to $d$ which is the number of spatial coordinates
and $\mbox{\boldmath$\hat{p}$}^2 \equiv \sum_{i=1}^d (\hat{p}_i)^2$.  
Hereinafter, we use index $i,j$ for spatial coordinates
and $a,b$ for all spacetime coordinates. 
Jacobi identity determines the full algebra:
\begin{eqnarray}
 \left[\hat{x}^i,\hat{x}^j\right]&=&-2i\hbar \beta(1+\beta
  \mbox{\boldmath $\hat{p}$}^2)\hat{L}^{ij}. \label{GUP3} \\
\left[\hat{p}^i,\hat{p}^j\right]&=&0. \label{GUP4}
\end{eqnarray}
Here $\hat{L}^{ij}$ are angular momentum like operators 
$\hat{L}^{ij}\equiv \frac{1}{2(1+\beta \mbox{\boldmath
$\hat{p}$}^2)}(\hat{x}^i\hat{p}^j-\hat{x}^j\hat{p}^i+\hat{p}^j\hat{x}^i-\hat{p}^i\hat{x}^j)$. 
Because operators $\hat{p}^i$ commute with each other, we construct a
theory in momentum space representation. In momentum space
representation, momentum operators are diagonalized simultaneously and
we do not distinguish eigenvalues of momentum $p_i$ from operators
$\hat{p}_i$. In the following, we set Planck
constant $\hbar$ to be 1 for simplicity. 

Lagrangian in $d+1$ dimensional spacetime \cite{Matsuo:2005fb} is   
\begin{eqnarray}
{\cal L}&=& -\frac{1}{2}\int^{\infty}_{-\infty}d^dp(1+\beta
    \mbox{\boldmath$p$}^2)^{-1}\phi(-p,t)
    \left[\partial_t^2+\mbox{\boldmath$p$}^2+m^2\right]
\phi(p,t), \label{scalaraction} \\
\mbox{where, \boldmath$p$}^2&\equiv&\sum_{i=1}^d (p_i)^2. \nonumber
\end{eqnarray}
The difference from ordinary quantum field theory is a prefactor
$(1+\beta \mbox{\boldmath$p$}^2)^{-1}$ in Lagrangian. 
Using the Bjorken-Johnson-Low prescription\cite{Bjorken:1966jh}, from behavior 
of $\mbox{T}^{\ast}$-product between $\phi(p,t)$ and $\phi(p',t')$, we
obtain canonical commutation relation: 
\begin{eqnarray}
\left[ \phi(p,t), \partial_t \phi(p',t)\right]=i
(1+\beta \mbox{\boldmath$p$}^2)\delta^d(p+p'). \label{Lcomm1}
\end{eqnarray}

As we can see from this equation, a deforming prefactor 
$(1+\beta \mbox{\boldmath $\hat{p}$}^2)$ of Heisenberg algebra in
the first quantization (\ref{GUP2}) also appears in canonical commutation
relation of the second quantized field theory. 

In a fermion field case, we encounter a difficulty at constructing the
second quantized Hilbert space which does not appear in a scalar
system. 
Note that a system of spin 0 particles contains only spin 0 particle. 
By contrast, a system of spin $\frac{1}{2}$ particles is not
closed with only fermions in the sense that it contains bosons as bound states. Therefore  
algebra of fermion fields must be introduced to be consistent 
with that of bosons fields. Because the scalar fields in our theory have a different commutation
relation (\ref{Lcomm1}) from ordinary one, we must construct fermion
fields so that the composite fields which correspond to scalar
particles have the same commutation relations. Or, in two-dimensional ordinary 
quantum field theory we could use the concepts of bosonization and
fermionization which associate fermion fields with boson fields. 
However, it is obscure which of these principles which relate bosons and
fermions remains unchanged in GUP or in 
extremely high energy regions. 
Instead of handling this problem directly, we use
supersymmetry to construct quantized field theory of fermion. 
This is because string theory accommodates this symmetry and therefore it is
expected that this symmetry is reflected in GUP or in 
extremely high energy regions. 

In the next section we construct a quantum field theory of fermions
which is consistent with the above scalar theory by using supersymmetry.  

\section{Supersymmetry in GUP}
\label{sec:susy}
In two and three-dimensional spacetime, 
a system with a real scalar and a Majorana fermion has 
a special symmetry between a boson and a fermion, namely 
supersymmetry. Thus we construct a quantum field theory of
fermion in GUP to have a similar symmetry between bosons and fermions
with an above-reviewed scalar
system in two and three-dimensional spacetime.

Our notation for two and three-dimensional
spacetime is as follows: In those
dimensional spacetime (with signature $- +$ or $- + +$) the Lorentz group has 
a real (Majorana) two-component spinor representation $\psi^{\alpha}$. 
In the following, we explain the notation of three-dimensional
spacetime. Reduction to two-dimensional spacetime is trivial.  
We define a representation of Gamma matrices by Pauli matrices\footnote{
Pauli matrices are $\sigma_1=\left(
\begin{array}{cc}
0 & 1 \\
1 & 0
\end{array}
\right),\sigma_2=\left(
\begin{array}{cc}
0 & -i \\
i & 0
\end{array}
\right),\sigma_3=\left(
\begin{array}{cc}
1 & 0 \\
0 & -1
\end{array}
\right).$
} as follows:
\begin{eqnarray}
\{\Gamma^a,\Gamma^b\}&=& 2\eta^{ab}=2\mbox{diag}(-++), \\
\Gamma^0&=&-i\sigma_2,\Gamma^1=\sigma_1, \Gamma^2=-\sigma_3.
\end{eqnarray}
Spinor indices are lowered and raised by charge conjugation matrix 
$C_{\alpha\beta}\equiv \Gamma^0$ and its inverse matrix $C^{-1}$: 
\begin{eqnarray}
\psi_{\alpha}&=& \psi^{\beta}C_{\beta \alpha}(=\bar{\psi}_{\alpha}),\psi^{\alpha}=
 \psi_{\beta}(C^{-1})^{\beta \alpha}.
\end{eqnarray}

Because the algebra of scalar field (\ref{Lcomm1}) is deformed from
usual one, it is natural to expect that  
supersymmetry algebra may also be deformed from ordinary one. 
We generalize supersymmetry algebra and its actions on a scalar
field $\phi$, a Majorana fermion $\psi$ and an auxiliary field $F$ 
with parameter $\epsilon^{\alpha}$as follows:
\begin{eqnarray}
\left[\bar{\epsilon}_1 \hat{Q},\bar{\epsilon}_2 \hat{Q} \right]
 &=&2\Delta \bar{\epsilon}_1\Gamma^a\epsilon_2\hat{P}_a, \label{susyalg}\\
\delta \phi (p,t) &=& i\bar{\epsilon}\psi(p,t), \label{susyphi}\\
\delta \psi^{\alpha}(p,t) &=& A_1 F(p,t)\epsilon^{\alpha}- A_2 
\{(\bar{\epsilon}\Gamma^0C^{-1})^{\alpha}\partial_t+(\bar{\epsilon}\Gamma^jC^{-1})^{\alpha}(ip_j)\}\phi(p,t), \label{susypsi}\\
\delta F(p,t) &=& A_3 i \bar{\epsilon}(\Gamma^0\partial_t+\Gamma^j(ip_j))\psi(p,t). \label{susyF}
\end{eqnarray}
Here, we introduce factors $\Delta, A_i$ as functions of a deforming parameter 
$\beta$ and momentum. These factors should reduce to 1 in the limit of 
$\beta \to 0$ and will be determined later by consistency
conditions.

From the closeness of algebra on each fields, we obtain conditions 
\begin{eqnarray}
A_1A_3=A_2=\Delta. \label{cond1}
\end{eqnarray}
We also generalize a Lagrangian by introducing factors $B_i$, which
are functions of a deforming parameter 
$\beta$ and momentum and are to be determined as well:  
\begin{eqnarray}
{\cal L}=\int dp^d && \left\{ -\frac{B_1}{2}\phi(-p,t)(\partial_t^2+\mbox{\boldmath$p$}^2)\phi(p,t)
-\frac{i B_2}{2}\bar{\psi}(-p,t)(\Gamma^0\partial_t+(ip_i)\Gamma^i+m)\psi(p,t)\right. \nonumber \\
&&\left.
+\sqrt{B_1B_3}m\phi(-p,t)F(p,t)+\frac{B_3}{2}F(-p,t)F(p,t)\right\}. \label{3daction1}
\end{eqnarray}
Here $d$ is the number of spatial coordinates (1 or 2).
By integrating out the field $F$, we obtain Lagrangian with the scalar field
and the Majorana field:
\begin{eqnarray}
{\cal L}=\int dp^d && \left\{ 
		    \frac{B_1}{2}\phi(-p,t)(\partial_t^2+\mbox{\boldmath$p$}^2+m^2)\phi(p,t) \right. \nonumber \\
&&- \left. \frac{i B_2}{2}\bar{\psi}(-p,t)(\Gamma^0\partial_t+(ip_i)\Gamma^i+m)\psi(p,t)\right\}. \label{3daction2}
\end{eqnarray}
Invariance of Lagrangian (\ref{3daction1}) under supersymmetry variations  
(\ref{susyphi})-(\ref{susyF}) leads following conditions;
\begin{eqnarray}
A_1B_2&=&\sqrt{B_1B_3}, \nonumber \\
A_1B_2&=&A_3B_3, \nonumber \\
B_1&=&A_1A_3B_2. \label{cond2}
\end{eqnarray}
From conditions (\ref{cond1}) and (\ref{cond2}), only $B_1$ and $B_2$
remain to be determined. (Factor $A_1$ can be absorbed into
normalization of a field $F$ and we set it to be 1 for a field $F$ to be
an auxiliary field.) 
Noether's current for supersymmetry can be calculated from Lagrangian
(\ref{3daction2}) and supersymmetry charge is found to be 
\begin{eqnarray}
Q^{\alpha}=\int dtdp^d B_1\{&-&\psi^{\alpha}(-p,t)\partial_t\phi(p,t)+(\Gamma^i\Gamma^0 
            \psi(-p,t))^{\alpha}(ip_i)\phi(p,t) \nonumber \\
&+&m(\Gamma^0\psi(-p,t))^{\alpha}\phi(p,t)\}.
\end{eqnarray}
Then, we obtain Hamiltonian of this system from supersymmetry charge and
algebra (\ref{susyalg}), 
\begin{eqnarray}
{\cal H}&=&P^0=-\frac{1}{4}\frac{B_2}{B_1}(C\Gamma^0)_{\alpha\beta}\{Q^{\alpha} ,
 Q^{\beta}\}.
\end{eqnarray}
Using the Bjorken-Johnson-Low prescription, from behaviors  
of $\mbox{T}^{\ast}$-product between fields, we
obtain canonical commutation relations as follows,
\begin{eqnarray}
\left[\phi(p,t),\partial_t \phi(q,t)\right]&=&\frac{i}{B_1}\delta(p+q), \\
\left\{ \psi^{\alpha}(p,t),\psi^{\beta}(q,t) \right\} &=&
-\frac{(\Gamma^0C^{-1})^{\alpha\beta}}{B_2}\delta(p+q).
\end{eqnarray}
Thus we can write the Hamiltonian in the following form;
\begin{eqnarray}
{\cal H}=\int dp^d && \frac{B_1}
{2}\{\pi(-p,t)\pi(p,t)+\phi(-p,t)(\mbox{\boldmath$p$}^2+m^2)\phi(p,t) \} \nonumber \\
&&
+\frac{i B_2}{2}\bar{\psi}(-p,t)((ip_i)\Gamma^i+m)\psi(p,t).
\end{eqnarray}
Here, we use conjugate momentum $\pi(p,t)=\partial_t\phi(-p,t)$ and indices 
$i$ runs from 1 to $d$. 

There is another condition which can be used to determine the factors $B_1$ and $B_2$. It comes from
the free energy of supersymmetric vacuum. From algebra (\ref{susyalg}),
supersymmetric state has zero energy:
\begin{eqnarray}
0&=&\frac{1}{2}\Trb \ln(B_1(E^2+\mbox{\boldmath $p$}^2+m^2))-\frac{1}{4}\Trf
 \ln(B_2^2(E^2+\mbox{\boldmath $p$}^2+m^2)).
\end{eqnarray}
This fact leads to the condition;
\begin{eqnarray}
 B_1&=&B_2^2.
\end{eqnarray}
Here $\Trb$ and $\Trf$ represent trace in bosonic and fermionic Hilbert
space respectively. 

Lastly, we set $B_1=(1+\beta \mbox{\boldmath $p$}^2)^{-1}$ as
we can see from the scalar action (\ref{scalaraction}). This determines all of
the introduced factors as follows; 
\begin{eqnarray}
\Delta=A_2=A_3=B_2 &=& (1+\beta \mbox{\boldmath $p$}^2)^{-\frac{1}{2}}, \\
A_1=B_3&=& 1, \\
B_1&=&(1+\beta \mbox{\boldmath $p$}^2)^{-1}.
\end{eqnarray}
Thus we construct quantized fields of fermion which is consistent with
scalar fields (\ref{Lcomm1}) as 
\begin{eqnarray}
\left\{ \psi^{\alpha}(p,t),\psi^{\beta}(q,t) \right\} &=&
-(1+\beta \mbox{\boldmath $p$}^2)^{\frac{1}{2}}(\Gamma^0C^{-1})^{\alpha\beta}\delta(p+q).
\end{eqnarray}

Note that a factor $\Delta$ is not equal to 1
no matter how we set $A_1$.
Therefore this supersymmetry algebra is deformed from an usual one as  
\begin{eqnarray}
\left[\bar{\epsilon}_1 Q,\bar{\epsilon}_2 Q \right]
 &=&2(1+\beta \mbox{\boldmath $p$}^2)^{-\frac{1}{2}} \bar{\epsilon}_1\Gamma^a\epsilon_2P_a.
\end{eqnarray}

There is no difficulty in generalizing the above quantum fields of
fermion to higher $d+1$ dimensions than three dimensions. The action is as follows: 
\begin{eqnarray}
{\cal L}=\int dp^d && \left\{ 
- \frac{i}{(1+\beta \mbox{\boldmath $p$}^2)^{\frac{1}{2}}}\bar{\psi}(-p,t)(\Gamma^0\partial_t+(ip_i)\Gamma^i+m)\psi(p,t)\right\}. \label{fermionaction}
\end{eqnarray}
There appears a universal prefactor 
$(1+\beta \mbox{\boldmath $p$}^2)^{-\frac{1}{2}}$ comparing with usual fermion
action regardless as to whether there were supersymmetry or not. This
prefactor ensures that fermion fields are compatible with the scalar
fields which had been constructed in our previous paper \cite{Matsuo:2005fb}.   

From the actions (\ref{scalaraction}) and (\ref{fermionaction}),   
we also have supersymmetric field theory in four dimensions 
with an complex scalar and a Majorana (or Weyl) fermion just as
a corresponding ordinary 
field theory has supersymmetry in four dimensions.   
\section{Conclusion and Discussions}
\label{sec:summary}
In summary, we have constructed a quantum theory of free fermion field based on 
the deformed Heisenberg algebra. It is consistent with already
proposed scalar theory through supersymmetry. We start with a system
with an real scalar and a Majorana fermion 
in two- and three-dimensional spacetime and determine supersymmetric
action. We found that supersymmetry
algebra is deformed from an usual one. An extension to higher dimensions
are trivial and there is also supersymmetric theory in four-dimensional
spacetime. 

We conclude with a brief discussion on Lorentz invariant extension of
our theory. 
Lorentz invariant extension of deformed Heisenberg algebra (\ref{GUP2}) 
is known as a sort of `doubly special relativity' or `$\kappa$-deformation' (for
example, see \cite{Lukierski:2002df}, \cite{Amelino-Camelia:2002vy} and
references therein), 
\begin{eqnarray}
 \left[\hat{x}^a,\hat{p}_b\right]&=&i\hbar (1+\beta \hat{p}^2)\delta^a_{\sp b}. \label{GUP5}
\end{eqnarray}
Here $a,b$ run from 0 to $d$ 
and $\hat{p}^2 \equiv -(\hat{p}^0)^2+\sum_{i=0}^d (\hat{p}_i)^2$.
Thus we claim that an action where the factor 
$(1+\beta \mbox{\boldmath $p$}^2)$ is replaced with a new factor 
$(1+\beta p^2)$ describes quantum field theory of doubly special
relativity. 
In such case, time slice is not well-defined because of the existence of 
minimal time interval. Therefore there is no canonical formalism.

\begin{acknowledgments}
The author is grateful to 
T.~Matsuo, 
K.~Oda, 
T.~Tada,   
and 
N.~Yokoi 
for valuable discussions. 
The author is supported by 
the Special Postdoctoral Researchers Program at RIKEN. 
\end{acknowledgments}



\end{document}